\documentclass[aps,prl,twocolumn,showpacs,superscriptaddress]{revtex4}

\usepackage{bm}
\usepackage[latin1] {inputenc}
\usepackage{graphicx}
\begin{document}

\title{Soft
{\em Listeria}
: actin-based propulsion of liquid drops}
\author{Hakim Boukellal, Otger Campàs, Jean-Fran\c cois Joanny, Jacques Prost and Cécile Sykes}
\affiliation{Institut Curie, UMR 168, 26 rue d'Ulm, F-75248 Paris Cedex
05, France}
\date{\today }

\begin{abstract}
  We study the motion of oil drops propelled by actin polymerization in cell extracts. 
Drops deform and acquire a pear-like shape 
under the action of the elastic stresses
exerted by the actin comet. We solve 
this free boundary problem and calculate the drop shape taking into account the elasticity of the actin gel and the variation 
of the polymerization velocity with  normal stress. The pressure balance on the liquid drop imposes a zero 
propulsive force if gradients in 
surface tension or internal pressure are not taken into account. 
Quantitative parameters of actin polymerization are obtained by fitting theory to experiment.

\end{abstract}

\maketitle

Actin polymerization is a key element in the motility of most cells and bacteria. The bacteria 
{\em Listeria monocytogenes} 
are propelled inside cells by the growth of a soft elastic comet made of a filamentous actin network. Actin polymerizes 
at the back of the advancing 
bacterium. 
The biochemistry of the comet formation is now well-understood \cite{pollard}.  Theoretical approaches have 
been proposed to explain the physical mechanism of force production. They differ by the scale at which they 
describe the mechanism. Molecular models by 
 Mogilner et al \cite{Mogilner} consider the Brownian flexibility of growing 
actin filaments, whereas Carlsson \cite{Carlsson} concentrates on the effect of branching and growth of the actin 
network. Gerbal et al \cite{Gerbal} analyze at a mesoscopic scale 
the elastic stresses exerted by 
the deformed comet gel on the bacterium resulting in the propulsive force \cite{prost}. 

A further step in understanding the propulsion
mechanism is provided by the study of biomimetic experimental systems where {\em Listeria} are replaced by solid spherical 
 beads on which actin polymerization promoters \cite{theriotpnas,anne,Vincent,welch}
are attached. These beads mimic closely the natural propulsion mechanism of {\em Listeria} with comet tail formation, 
after the breaking of the initial spherical symmetry \cite{theriotncb,anne}.
Biomimetic systems allow for a systematic variation of the parameters and thus for a quantitative comparison to theory.

The aim of this letter is to demonstrate both theoretically and experimentally the
mechanism of force production due to the elastic stresses exerted by the actin comet gel.
A newly designed experimental system is made of oil drops with actin polymerization promoters attached on their surface. 
Once placed in cell extracts, such an oil drop moves 
by actin polymerization and deforms under the action of the elastic stresses exerted by the gel.
The same squeezing effect is observed on endosomes \cite{Tauton} driven 
by actin comets and synthetic vesicles covered with the bacterial protein ActA  \cite{VanOudenaarden, giardini}. However, for
liquid drops, the knowledge of the surface tension and the constant volume allow for a quantitative analysis of the observed shape
and thus the determination of the elastic stress distribution on the drop surface. 

An emulsion of oil drops is obtained by sonicating ($5$ seconds, $90$ W) 
a mixture of edible oil, Isio4, and buffer ($2.3 \%$ 
oil in borate buffer $100$mM, pH $8.5$). The actin polymerization promoter VCA
is derived from the Wiskott-Aldrich syndrom protein  (WASP) 
and purified as described in Fradelizi et al \cite{Fradelizi}. VCA is adsorbed onto the oil drops by 
incubating 10 $\mu$l of the emulsion with a $0.2$mg/ml VCA solution in borate buffer. A volume of $0.2$ $\mu$l 
of the emulsion coated with VCA is added to $15$ $\mu$l of HeLa extracts 
prepared as explained in Noireaux et al \cite{Vincent} with a  final 
protein concentration of $19$mg/ml. As usual the extracts are supplemented with 0.019mg/ml 
G-actin, 0.06mg/ml rhodamin actin, 1mM ATP, 27mM creatine phosphate, 1mM DTT. 

\begin{figure}
{\centering\resizebox*{6cm}{!} {\includegraphics{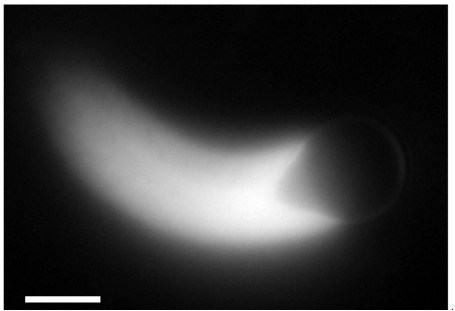}} }
\caption{\label{fig: figure1}
Oil drop covered with VCA placed in HeLa cell extracts supplemented with 
Alexa-actin and observed by fluorescence microscopy. 
The actin comet appears bright. Bar 4 $\mu$m.}
\end{figure}

\begin{figure}
{\centering\resizebox*{6cm}{!} {\includegraphics{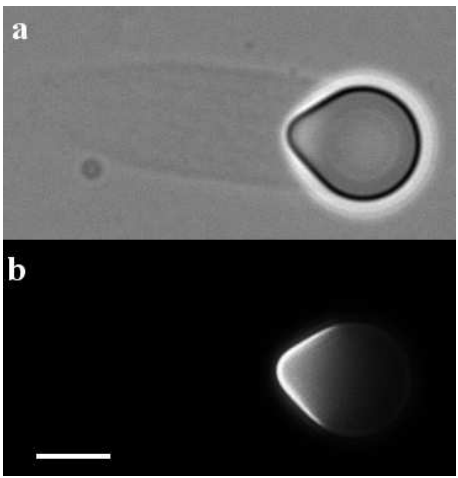}} }
\caption{\label{fig: WA}
Oil drop covered with FITC-labeled VCA  and placed in HeLa cell extracts observed (a) by bright field microscopy and 
(b) by fluorescence microscopy. 
The VCA appears bright. Bar 3 $\mu$m.}
\end{figure}

The sample is 
observed by bright field or fluorescence microscopy using an Olympus BX51(Germany)and 
Metamorph software (Princeton Instruments, USA).
The drop radii range from $1.5$ to $5.5$ $\mu$m.
As with {\em Listeria},
actin polymerizes only at the interface between the drop and the comet and depolymerizes at 
the back of the comet. When placed in
extracts, actin is first polymerized 
on the drops with a spherical symmetry. After roughly one hour, symmetry is broken for c.a. $70\%$ 
of the drops. They develop an actin comet, deform into a pear-like shape and move. 
Smaller drops are less deformed than larger drops.
An example of a moving drop is shown on Fig.\ref{fig: figure1}.
At the beginning of the experiment, the actin polymerization factor VCA is uniformly distributed 
around the spherical drop. After deformation of the
drop and formation of the comet, fluorescence intensity measurements using FITC-labeled VCA  \cite{fitc}
show that $90\%$ of the VCA is found on the 
comet side of the drop, as seen on Fig.\ref{fig: WA}.
This means either that VCA has been displaced from the interface to the bulk of the extract except where the gel is present,
or that all VCA has been collected by the gel during the symmetry breaking process. In any event this also means 
that the VCA surface density is comparable to the density of filament extremities at the surface. The average distance $d$ between 
filament extremities is larger than the close-packing  distance. A lower bound of $d$ is $10$nm. The surface 
tension change due to the 
presence of VCA at the interface is then of order $kT/d^2 < 4{\textrm x}10^{-2}$mN/m, more than $100$ times smaller than 
the oil-extract
tension that we measure by the pendant drop method  $\gamma_0=4\pm 0.6$ mN/m.

\begin{figure}
{\centering\resizebox*{6cm}{!} {\includegraphics{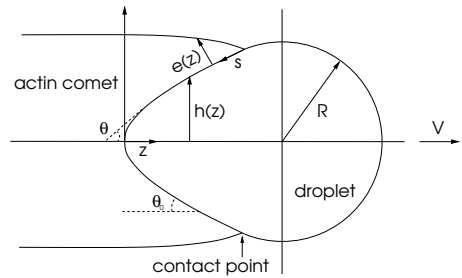}} }
\caption{\label{fig: figure2}
Sketch of the moving oil drop. For theoretical analysis, the shape is parameterized 
by the local thickness $h$ and the local angle of the tangent $\theta$}.
\end{figure}

The experimental shape of the drop is sketched on Fig.\ref{fig: figure2}.
The front part is a spherical cap of radius 
$R$ not covered by actin. 
The radius $R$ is different from the radius 
of the undeformed spherical drop $R_0$ 
that fixes the volume $4\pi R_0^3 /3$. The back part of the drop, surrounded by the comet, 
has a blunted cone-like shape (rotationally symmetric around the direction of motion). 
In order to give a theoretical description of the drop shape, 
we parameterize it by the liquid thickness $h$ and the 
local angle $\theta$ between the tangent to the shape and the direction of motion. The spherical cap and the cone
match at the triple line 
between the drop, the comet and the surrounding solvent; we call the tangent angle at the contact point 
$\theta_{m}$ where the corresponding oil thickness is $h_{m}=R \cos \theta_{m}$. The pressure inside
the drop varies from point to point since the drop motion induces an internal flow. At the interface with the extract, 
in the spherical part, it is given by Laplace's law $P_{in}=2\gamma_0 /R + P_0$, 
where $P_0$ is the pressure in the surrounding liquid. At any point inside the drop, the pressure  differs from $P_{in}$ and reads
$P=P_{in}+\delta P$.
At a point of thickness $h$ the local stress balance along the normal of the drop
is given by the local Laplace's law
\begin{equation}
\label{force}
\frac {2\gamma_0} R+ \delta P(h)= \gamma \left(\frac {\cos \theta(h)} h + \frac  {d\cos \theta(h)} {dh} \right) -\sigma_{nn}(h)
\end{equation}
where $\sigma_{nn}$ is the normal stress at the surface of the comet, exerted by the drop 
($\sigma_{nn}$ is positive, dilative stress, if the comet pulls on the drop and negative, compressive stress,
 if the comet pushes the drop). Although we argued that the surface tension gradients are small, we consider here that the 
surface tension $\gamma=\gamma_0+\delta\gamma(h)$ varies along the interface; it is constant in the spherical part 
with a value $\gamma_0$ and it is continuous at the contact line.
The total elastic force $F_e$ exerted by the comet on the drop 
is obtained by integrating the projection of the normal elastic stress on the direction of motion. 
\begin{equation}
F_e/2\pi= \int_0^{h_{m}} dh \ h\left(\cos \theta \frac{d \delta \gamma} {dh} +\delta P \right)
\label{propforce}
\end{equation}
If we ignore both the pressure gradient inside 
the drop and the surface tension gradient, the propulsive force equals zero. 
This result holds for any axisymmetric drop shape whatever the 
elastic stress distribution along the surface, independent of any model
for the comet elasticity and the actin polymerization velocity.
This hypothesis, used in references 
\cite{VanOudenaarden, giardini}, is unable to produce an estimate of the propulsive force. 
The important consequence of Eq.\ref{propforce} is that an experimental measurement of 
the propulsive force must take into account 
the surface tension gradient and the flow inside the drop as discussed later.

We now proceed with the determination of the drop shape and the stress distribution. Given that the
surface tension variation is small along the drop contour and neglecting hydrodynamic effects,  
we consider, for local equations, that the surface 
tension and the internal pressure are constant.
The elastic stresses in the 
gel influence the polymerization kinetics; polymerization is normal to the surface of the drop and it 
is accelerated by a dilative stress and slowed 
down by a compressive stress. Classical rate theories \cite{kramers} predict a polymerization velocity 
$v_p$ varying as a Boltzmann law
\begin{equation}
\label{polymer}
v_p(h)=v_p^0 \exp [\sigma_{nn}(h) /\sigma_0], \qquad \sigma_0 \equiv kT/ {a^2 \delta}
\end{equation}
where $a$ is the distance between actin polymerization promoters on the drop surface, $\delta$ 
is of the order of the size of an actin monomer and $v_p^0$ is 
the polymerization
velocity in the absence of stress. 

The last equation determining the shape of the drop is the conservation of the gel volume upon polymerization. 
In a first approximation, we assume both that the gel density is constant and that the comet is a perfect cylinder. 
In a steady state,  the drop advances at a constant
velocity $V$. 
The local gel thickness $e$ shown on Fig.\ref{fig: figure2} is then such that $de/ds =\tan \theta$ where $s$ is 
the length along the drop contour. 
With these simplifying approximations, the local polymerization velocity is related to the advancing velocity by 
$v_p(h)=V \sin \theta(h)$.

This approximation does not allow the determination of the drop velocity $V$ since the propulsive force vanishes. 
We thus find a family 
of solutions for the drop shape parametrized by the advancing velocity. We 
determine the drop shape using the measured  
advancing velocity $V=0.15 \pm 0.03 \mu m$/min.

The shape of the comet-drop interface departs from a pure cone
when the Boltzmann factor is significantly larger than 1. This defines the size of the blunted region as  
$\ell \equiv \gamma a^2 \delta /kT$ which leads to $\sigma_0 =\gamma / \ell$.
In the following we consider $\epsilon\equiv \ell/R$ as a small number.

For $h$ of order $R$, the elastic stress $\sigma_{nn}(h)$ is small and can be neglected. Then Eq.\ref{polymer}
describes a perfect cone with 
angle $\theta=\theta_0$ such that  $V\sin \theta_0 =v_p^0$.
This result is independent of the polymerization law giving $v_p$ as a function of $\sigma_{nn}(h)$.
The measure of $\theta_0$ gives thus access to the polymerization velocity
in the absence of elastic stress. 

In the blunted region, $h$ is of order $\ell$ and we neglect the Laplace pressure term
on the left hand side of equation \ref{force}. In the vicinity of the rear point, the drop profile is given by
 $h^2=4\ell z / \log (1/ \sin \theta_0)$. 

The normal stress in the rear region of size $\ell$ is positive and of order $\gamma /\ell$. At the rear point, 
$\sigma_{nn}(h=0)=\gamma \log (1 / \sin \theta_0 )/ \ell $. 
In the conical region, the stress reads at lowest order in $\epsilon$
\begin{equation}
\label{stress}
\sigma_{nn}(h)=(\gamma /R) \left( -2+ R \cos \theta_0 /h \right)
\end{equation}
It is positive at the back of the drop (pulling the drop backwards) 
and negative in the front part of the cone 
(pushing the drop forwards) as qualitatively observed in reference 
\cite{VanOudenaarden}. It vanishes 
for a thickness $h=(R  \cos \theta_0) /2=h_{m}/2$.
This is in accordance with the prediction made in ref. \cite{Gerbal} 
that the actin gel could pull at the rear of {\em Listeria} 
and explains the pear-like deformation that we now describe quantitatively.
 
\begin{figure}
{\centering\resizebox*{8cm}{!} {\includegraphics{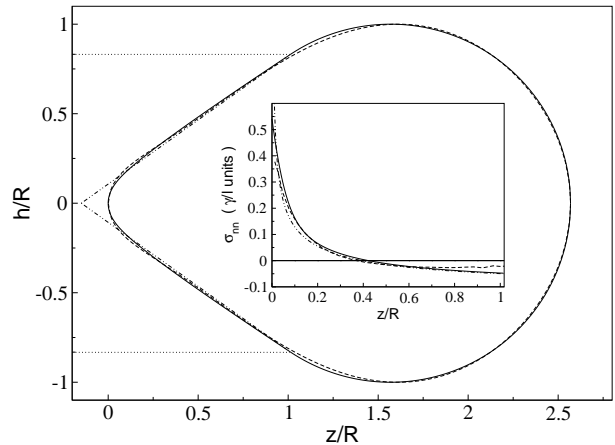}} }
\caption{\label{fig: figure3}
Calculated drop profile for $\epsilon =0.049$ and $\sin \theta_0=0.58$ (continuous line) compared to the experimental 
drop profile (dashed line) and to the zeroth order
asymptotic expansion (dash-dotted line). 
The insert  shows the normal stress distribution along the drop surface}
\end{figure}

A more detailed description of the drop profile is obtained by solving numerically equations \ref{force} and \ref{polymer}. 
We use the radius $R$ as a unit length; 
the drop profile depends on the two dimensionless parameters, $\epsilon$ and $\sin \theta_0=V/v_p^0$. 
On Fig.\ref{fig: figure3}, 
we show a comparison of the calculated 
and experimental profiles for the drop of Fig.\ref{fig: figure1}. The experimental profile
has been digitized and adjusted by a continuous curve. The best fitting parameters are 
$\epsilon=0.049$ and $\sin \theta_0=0.58$ ($\theta_0=35.6 ^{\circ}$); 
this gives a determination of the length $\ell=0.125 \mu m$, of the stress $\sigma_0=32 nN/\mu m^{2}$ 
and of the polymerization speed in the absence of stress $v_p^0 =1.4$nm/s \cite{footnote}.  
The local normal stress is obtained from the experimental
drop shape by using equation [\ref{force}] with a constant interfacial tension $\gamma_0$ and neglecting the variation of the 
 internal pressure $\delta P$.  The length $\ell$ being constant, smaller drops corresponding to larger $\epsilon$  
are less deformed, in agreement
with experiments.

In the description proposed so far, the curvature of the interface between the comet and the drop has a  
discontinuity at the triple line. 
In the spherical region, the curvature of the interface is $2/R$; in the conical region, the interface is curved 
only in one direction and the curvature is
$1/R$. The local forces normal to the interface are still balanced and the elastic normal stress in the comet at 
the triple line is $\sigma_{nn}(h_m)= -\gamma /R$.
(see insert of figure  \ref{fig: figure3}). 
At the
triple line, the gel thickness $e$ vanishes and cannot create a finite stress. This explains why on figure 
\ref{fig: figure3}, the theoretical and experimental normal stresses are in disagreement in the very vicinity of the triple line.
We now assume that the gel density remains constant 
but that the comet shape is
not a perfect cylinder in the region close to the triple line. Volume conservation then imposes a polymerization velocity 
$v_p =V \cos \theta \  de/ds$ where $s$ is the length along the interface.
When a gel element is created in a time $dt$, it is stretched by an amount $\delta u= \delta t (-V \sin \theta + v_p)$; 
the tensile stresses $\sigma_{ii}$ in the azimuthal direction 
($i=\phi$) and in the tangential direction along the interface ($i=t$) are then increased. At the level of scaling laws, 
we write this increase as $ \frac {d\sigma_{ii}}{ds}=\frac {E}{R_i V \cos \theta}(V \sin \theta - v_p)$ where $R_i$ is the radius 
of curvature of the interface in the direction $i$ and E an elastic modulus of the comet. 
In the vicinity of the triple line, we use 
the thin shell approximation and relate the tensile
and normal stresses in the gel  
$\sigma_{nn}=-e(\frac {\sigma_{tt}}{R_t}+\frac {\sigma_{\phi \phi}}{R_\phi})$. In this boundary layer, 
the two tensile stresses can be considered as constant; the matching to the pear-like shape imposes that 
the azimuthal tensile stress $\sigma_{\phi \phi}$ vanishes.
Defining the dimensionless
tensile stress ${\tilde \sigma_t} =\ell \sigma_{tt}/\gamma$, the normal stress in the boundary layer is then calculated as
\begin{equation}
\label{boundary}
\sigma_{nn}=- \frac{\gamma}{R} \frac{e{\tilde \sigma_t}}{\ell + e{\tilde \sigma_t}}
\end{equation}
As the thickness of the gel vanishes, the normal stress vanishes as expected. The boundary layer, where the comet is deformed, 
has a thickness $\ell / {\tilde \sigma_t}$ of order $\ell$ and is thus small
in the limit where $\epsilon$ is small. 
When $e$ is large, further away from the triple line, the comet reaches a cylindrical shape and the normal stress
is $-\gamma /R$; one can consider the comet as a perfect cylinder as done above. 

Our experimental observations on the shape of liquid drops propelled by actin polymerization are well described by the theoretical 
model based on a local normal force balance and on a Boltzmann variation of the polymerization velocity with normal 
stress. The results are robust if we use, for the polymerization velocity, the mathematical forms suggested by 
simulations on flat surfaces \cite{Carlsson}.
We demonstrate that the elastic propulsive force cannot be calculated from an 
elastic stress distribution that ignores both pressure variations 
inside the drop and surface tension gradients. 
We have estimated the surface tension gradient by assuming that the 
actin polymerization promoter density profile along the interface follows the gel elastic deformation. 
This leads to a propulsive force of order 
$F_e \sim 2\pi \gamma_0 \epsilon_G /E$ where $\epsilon_G$ is the Gibbs elastic modulus of the interface. 
With reasonable values of the parameters, we find a propulsive force of order $100$pN. This is however a lower bound
since there could exist other contributions to the surface tension gradient.
The advancing velocity of the drop results from a balance between 
the propulsive force and the friction force between the comet and the drop but
a precise study of the velocity selection will require a more refined analysis.

In our experiments, as well as in reference \cite{VanOudenaarden}, the motion stops when the drop 
(or the liposome) becomes spherical.
The large stress at the back of the drop
could lead to ``cavitation'' or rupture of the links between the drop and the comet. The elastic stresses 
exerted on the drop then relax and provoke the experimentally observed arrest. We expect the stress 
distribution to be similar for other types of actin propelled 
objects such as the bacteria {\em Listeria} or solid beads. This would explain the 
observation of hollow comets of {\em Listeria} \cite{Tauton}.
A final output of our analysis is an estimate of 
the polymerization velocity of an actin gel in the absence of stress and its variation with 
stress.

\begin{acknowledgements}
J.F.J. is grateful to A. van Oudenaarden for sending a copy of reference \cite{VanOudenaarden}. 
O.C. thanks the European Network PHYNECS and Ministerio de Educaci\'on, Cultura y Deporte
 for financial support. We thank J.Casademunt 
and K.Zeldovitch (Institut Curie) for help in the numerical work  and V.Noireaux for initiating the experiments.
\end{acknowledgements}


\begin{thebibliography}{99999}
\bibitem{fitc}For FITC (fluorescein isothiocyanate) labeling, VCA is incubated for one hour 
with $0.08$ mg/ml of FITC (in 100mM borate buffer, PH 8.5) 
and dialysed against PBS at $4$°C.
\bibitem{footnote} Series of measurements in a given preparation of cell extract show a dispersion both in $\ell$
and $v_p^0$ of order $\pm 20\%$.  A  much larger spread of several hundred percent is found 
between different preparations (leading to different total protein concentrations).
\end{thebibliography}
\end{document}